\documentclass[iop]{emulateapj}

\usepackage{graphicx}
\usepackage{graphics}
\usepackage{amsmath}
\usepackage{txfonts}
\usepackage{epsfig}
\usepackage{natbib}
\usepackage{times}
\usepackage{amssymb}
\usepackage{color}
\usepackage{longtable}
\usepackage{lineno}
\newcommand\hipp{{\it{Hipparcos}}}

\newcommand\nustar{{\it{NuSTAR}}}

\newcommand\numHMXB{115}
\newcommand\numOBA{283}

\shorttitle{Clustering of HMXBs and OBAs in the SMC}
\shortauthors{Bodaghee et al.}

\begin{document}

\title{Evidence for low kick velocities among High-Mass X-ray Binaries in the Small Magellanic Cloud from the Spatial Correlation Function}
%\title{}

\author{Arash Bodaghee\altaffilmark{1}}
\author{Vallia Antoniou\altaffilmark{2,3}}
\author{Andreas Zezas\altaffilmark{4}}
\author{John A. Tomsick\altaffilmark{5}}
\author{Zachary Jordan\altaffilmark{1}}
\author{Brenton Jackson\altaffilmark{1}}
\author{Ryan Agnew\altaffilmark{1}}
\author{Eric Frechette\altaffilmark{1}}
\author{Ann E. Hornschemeier\altaffilmark{6,7}}
\author{J\'{e}r\^{o}me Rodriguez\altaffilmark{8}}

\altaffiltext{1}{Dept. of Chemistry, Physics and Astronomy, Georgia College and State University, Milledgeville, GA 31061, USA}
\altaffiltext{2}{Dept. of Physics, Box 41051, Science Building, Texas Tech University, Lubbock, TX 79409, USA}
\altaffiltext{3}{Harvard-Smithsonian Center for Astrophysics, Cambridge, MA 02138, USA}
\altaffiltext{4}{Dept. of Physics, University of Crete, GR-71003 Heraklion, Greece}
\altaffiltext{5}{Space Sciences Laboratory, University of California, Berkeley, CA 94720, USA}
\altaffiltext{6}{NASA Goddard Space Flight Center, Greenbelt, MD 20771, USA}
\altaffiltext{7}{Dept. of Physics and Astronomy, Johns Hopkins University, Baltimore, MD 21218, USA}
\altaffiltext{8}{Lab. AIM, CEA/CNRS/Universit\'{e} Paris-Saclay, Universit\'{e} de Paris, CEA-Saclay F-91191 Gif-sur-Yvette, France}

\begin{abstract}
We present the two-point cross-correlation function between high-mass X-ray binaries (HMXBs) in the Small Magellanic Cloud (SMC) and their likely birthplaces (OB Associations: OBAs). This function compares the spatial correlation between the observed HMXB and OBA populations against mock catalogs in which the members are distributed randomly across the sky. A significant correlation ($\sim$15$\sigma$) is found for the HMXB and OBA populations when compared with a randomized catalog in which the OBAs are distributed uniformly over the SMC. A less significant correlation (4$\sigma$) is found for a randomized catalog of OBAs built with a bootstrap method. However, no significant correlation is detected when the randomized catalogs assume the form of a Gaussian ellipsoid or a distribution that reflects the star-formation history from 40\,Myr ago. Based on their observed distributions and assuming a range of migration timescales, we infer that the average value of the kick velocity inherited by an HMXB during the formation of its compact object is 2--34\,km\,s$^{-1}$. This is considerably less than the value obtained for their counterparts in the Milky Way hinting that the galactic environment affecting stellar evolution plays a role in setting the average kick velocity of HMXBs.

\end{abstract}

%%__________________________________________________________________


\section{Introduction}

A High-Mass X-ray Binary (HMXB) is a system in which a compact object (usually a neutron star; sometimes a black hole) accretes wind material shed by a massive ($M \gtrsim 10\,M_{\odot}$) stellar companion. According to the mass-age relation for high-mass stars \citep{Sch92}, around $10$\,Myr is thought to elapse between stellar birth and supernova. This implies that HMXBs are young systems that do not have enough time to migrate far away from their birthplaces, i.e. OB Associations (OBAs) which are loose collections of young, massive O-type and B-type stars. Thus, the observed locations of HMXBs serve to trace recently active sites of massive star formation in a galaxy \citep[e.g.,][]{gri02}. 

In a previous study, we assessed the degree of spatial correlation between the HMXB and OBA populations in the Milky Way \citep{bod12} by building the two-point spatial cross-correlation function \citep[][]{pee80,lan93}. We found that the spatial distributions of the two populations were closely related, as expected. More importantly, we were able to use this correlation to derive average distances or durations for the HMXBs as they migrated away from their birthplaces over the past few million years. The migration is thought to be due to a combination of effects including recoil due to anisotropic mass loss from the primary to the secondary \citep{bla61}, dynamical ejection and cluster outflows \citep{pov67,pfl10}, or natal kicks. In fact, this natal kick could allow the HMXB to reach a significant velocity if the supernova event that created the NS is asymmetrical \citep{shk70}. An example of an asymmetric supernova was seen by \nustar\ in SN1987A situated in the Large Magellanic Cloud \citep{bog15}, although this object did not ultimately lead to an HMXB. The energy available for kicks depends on a number of factors internal to the system such as the mass ratio of the binary, magnetic fields, etc., but also on external factors such as the metallicity that affect the entire population. 

Unfortunately, cases in which an HMXB can be linked back to its parent OBA, and its velocity measured, are rare and only possible for relatively nearby systems in the Milky Way \citep[e.g.,][]{mir04}. Thus, a statistical approach applied to these populations offers a way to estimate the average kick velocity of the sample, and how much this value depends on the galactic environment in which these systems reside.

%-----------------------------Figure Start------------------------------
\begin{figure*}[!t]
\begin{center}
\includegraphics[width=\textwidth]{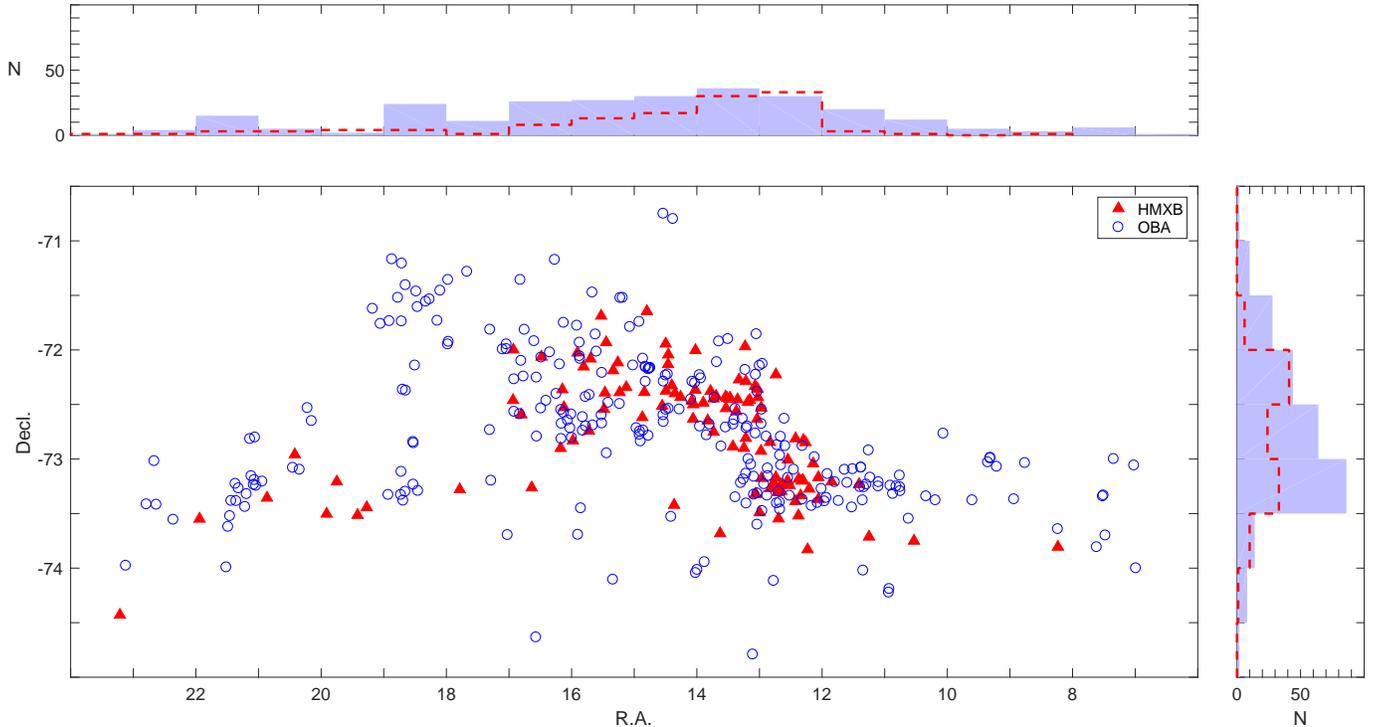}
\end{center}
\caption{Spatial distributions are presented for objects belonging to the Small Magellanic Cloud that are considered in this study. High-mass X-ray binaries are shown as red triangles \citep[\numHMXB\ HMXBs from][]{hab16}, while the locations of known OB Associations are represented by blue circles \citep[\numOBA\ OBAs from][]{bic08}. The histograms present the number of HMXBs (dashed red line) and observed OBAs (blue shaded region) over J2000.0 equatorial coordinates in degrees.}
\label{fig_sky}
\end{figure*}
%-----------------------------Figure End--------------------------------

At a distance of 61\,kpc \citep{hil05}, the Small Magellanic Cloud (SMC) is an irregular dwarf galaxy with a relatively large and active population of HMXBs \citep[e.g.,][]{liu05} that are bright, easy to resolve, and with low intervening extinction \citep{zar02} and photoelectric absorption \citep{kal05}. The SMC underwent chemical and stellar evolutionary processes that led to disparities in its HMXB population compared with those of the Milky Way \citep[e.g.,][]{cla78,maj04,dra06,leh10,lin10,min12,fra13,for20}. The primary difference is that stars in the SMC feature a lower average metallicity than stars in our galaxy \citep{luc98,ant16}. Another difference is that nearly all HMXBs in the SMC host Balmer-emission line stars \citep[so-called BEXBs:][]{coe05a,ant09,hab16}, while only two host supergiant OB stars \citep[so-called SGXBs:][]{mar14}. 

In the Milky Way, BEXBs are also the dominant subclass, but the percentage of SGXBs is an order of magnitude greater and continues to increase as new SGXBs are discovered \citep[][and references therein]{wal15}. This enhancement of HMXBs in the SMC (and BEXBs in particular) is attributed to a lower galactic metallicity \citep[e.g.,][]{lin10,kaa11,bas16} and a prolific episode of new star formation 40\,Myr ago \citep{dra06,ant10}. Remarkably, there are no confirmed HMXBs with black holes in the SMC, likely due to the preponderance of BEXBs \citep{zha04}. The SMC offers a homogeneous population of HXMBs that can be compared with an HMXB population in the Milky Way that evolved under a different star formation history.

The goal of the present work is to generate the two-point spatial correlation function for the HMXB and OBA populations in the SMC, as was done previously for these populations in the Milky Way. Section\,\ref{sec_data} presents the source populations and the analysis methods including the generation of randomized source distributions. In Section\,\ref{sec_res}, the main results are discussed then summarized.

%-----------------------------Figure Start------------------------------
\begin{figure*}[!t]
\begin{center}
\vspace{1mm}
\includegraphics[width=0.47\textwidth]{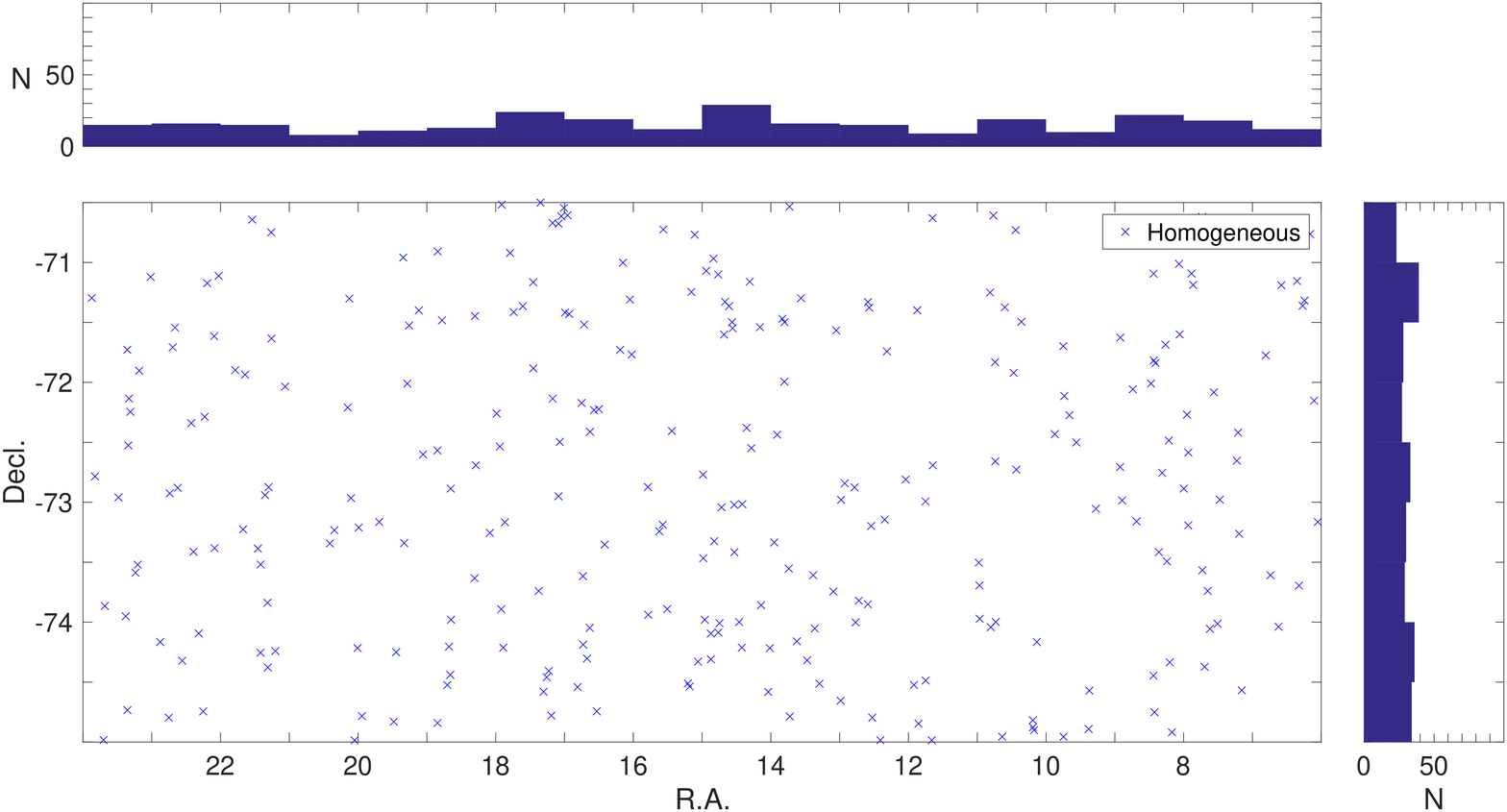}\hspace{5mm}\includegraphics[width=0.47\textwidth]{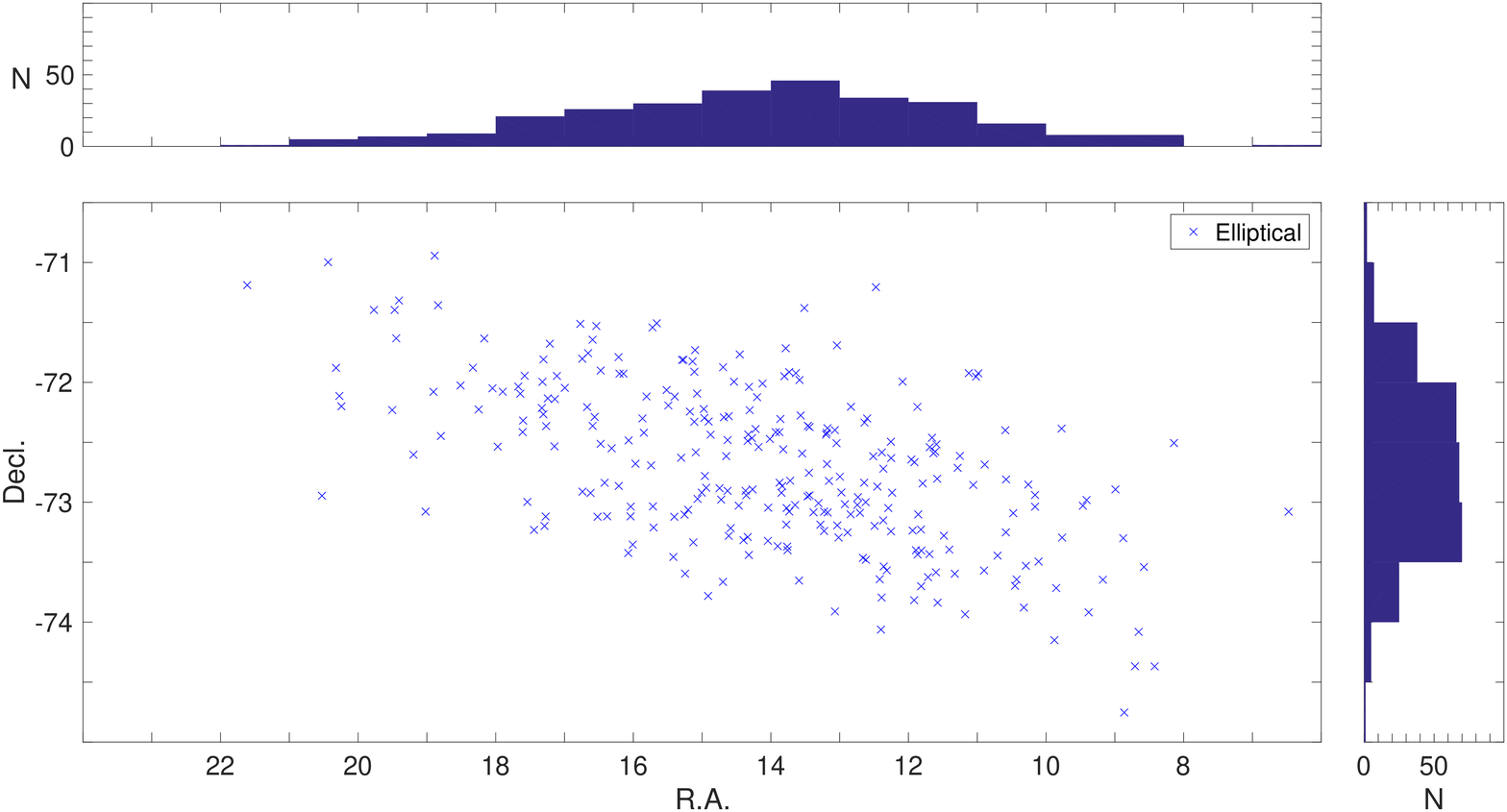}

\vspace{3mm}
\includegraphics[width=0.47\textwidth]{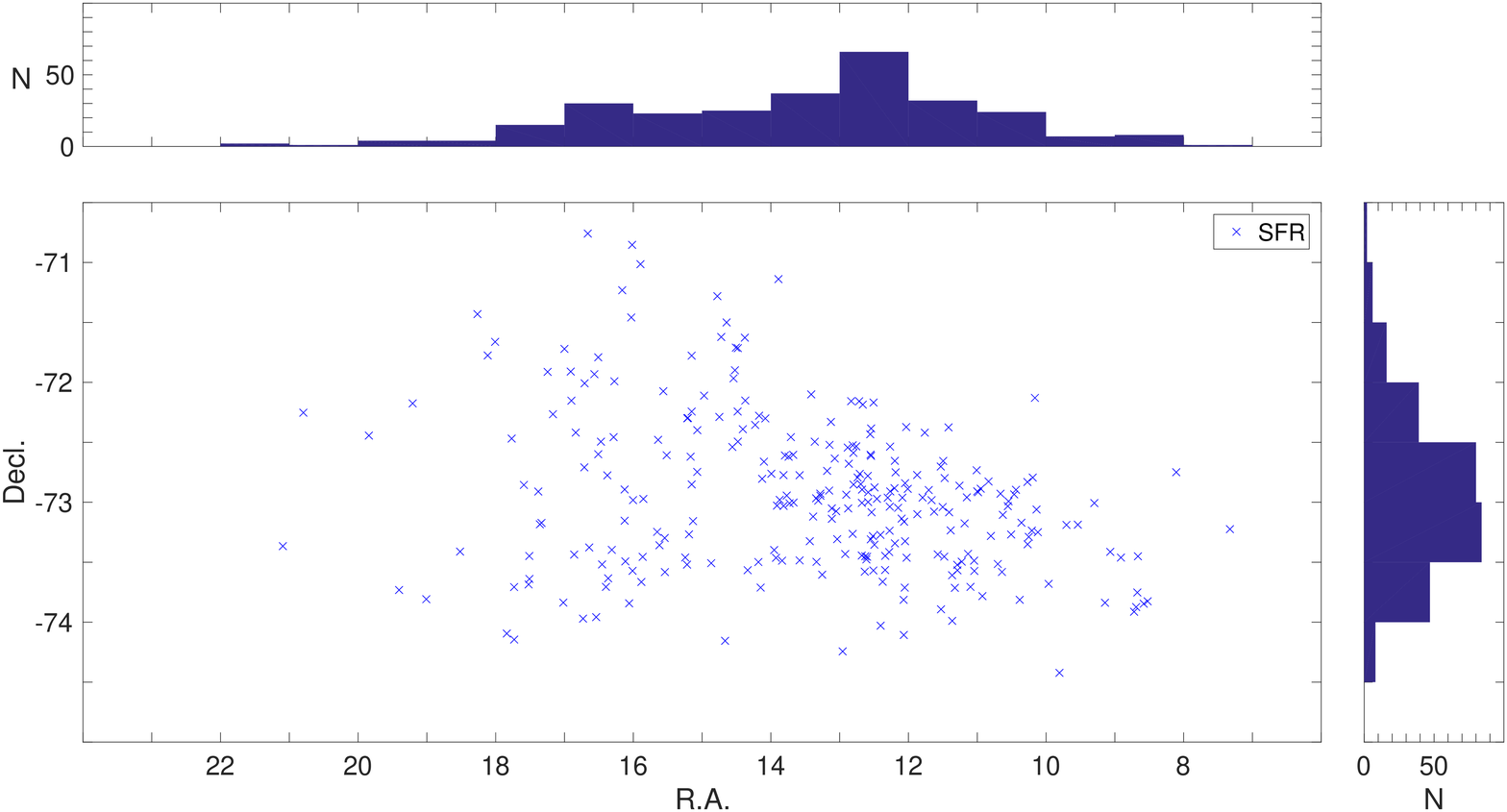}\hspace{5mm}\includegraphics[width=0.47\textwidth]{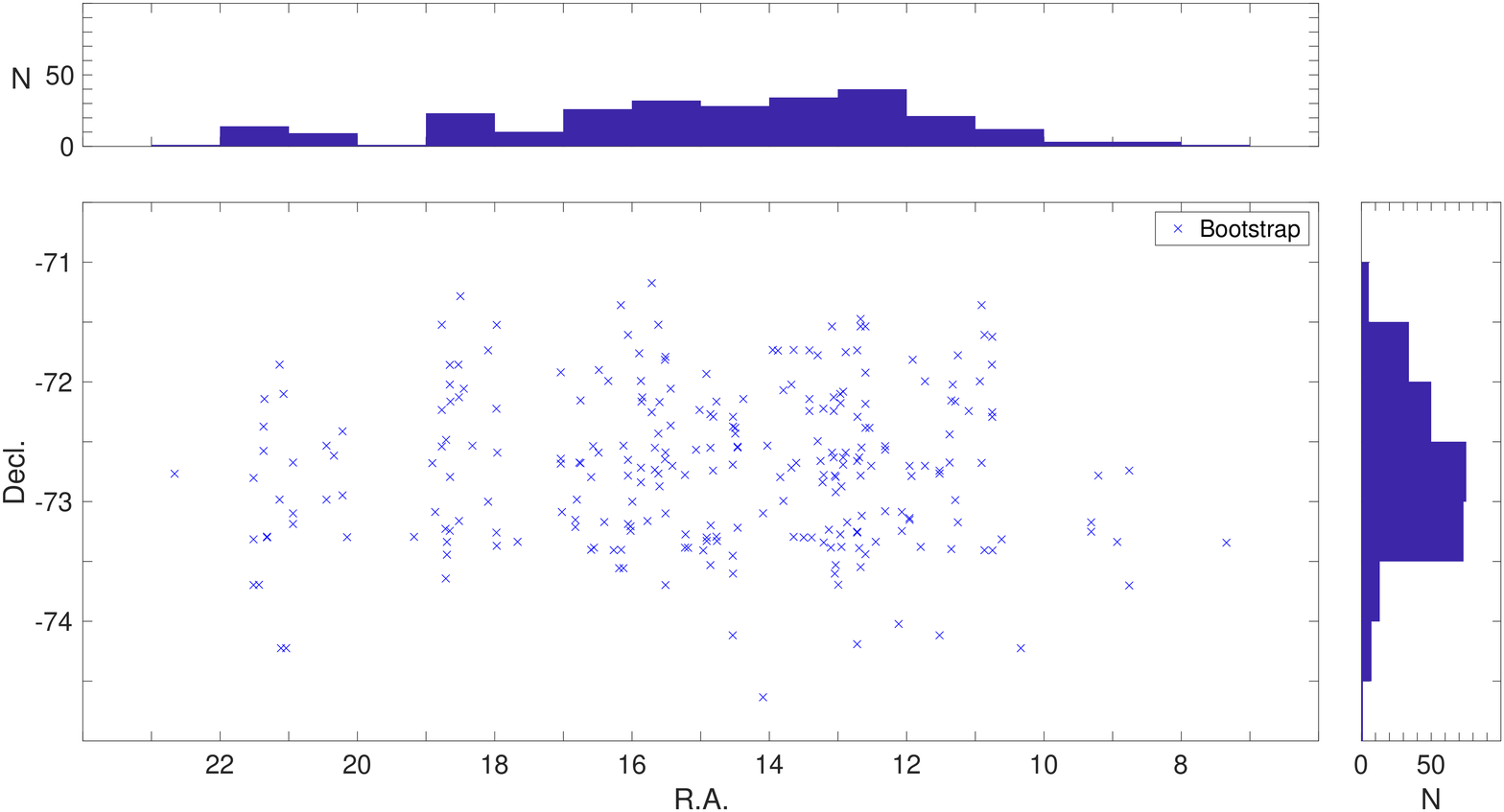}
\end{center}
\caption{Example subsets of the four randomized OBA catalogs used in this work. Each of the main panels presents the same sky area as shown in Fig.\,\ref{fig_sky}. Blue crosses represent \numOBA\ objects drawn at random (out of $10^{5}$ members) from each randomized OBA catalog; clockwise from the upper-left are Homogeneous, Elliptical, Bootstrap. and SFR. }
\label{fig_sky_random}
\end{figure*}
%-----------------------------Figure End--------------------------------

\section{Data \& Analysis Methods}
\label{sec_data}

\subsection{Observed and Randomized Catalogs}

Beginning with the list of 148 HMXB candidates in the catalog of \citet{hab16}, we exclude 27 objects whose HMXB classification is in doubt according to the catalog's authors. Of the remaining 121 confirmed or likely HMXBs, 6 objects are rejected due to being in the Magellanic Bridge. The range of R.A. for the final list of \numHMXB\ HMXBs considered in this study is 6$^{\circ}$--24$^{\circ}$. Within that range of R.A., there are \numOBA\ OBAs from the catalog of \citet{bic08}, which will be referred to as the ``observed OBA catalog.'' 

Figure\,\ref{fig_sky} presents the observed populations of HMXBs and OBAs as they are distributed within the boundaries of the sky region considered in this study. By eye, it is apparent that these populations are clustered together in space, as expected. The significance of any spatial clustering can be found by constructing the spatial correlation function. Essentially, this function takes the spatial distribution of the HMXB population and determines which provides a better match: the observed OBA catalog, or a mock OBA catalog in which the members are distributed randomly. 

%-----------------------------Figure Start------------------------------
\begin{figure*}[!t]
\begin{center}
\vspace{1mm}
\includegraphics[width=0.45\textwidth]{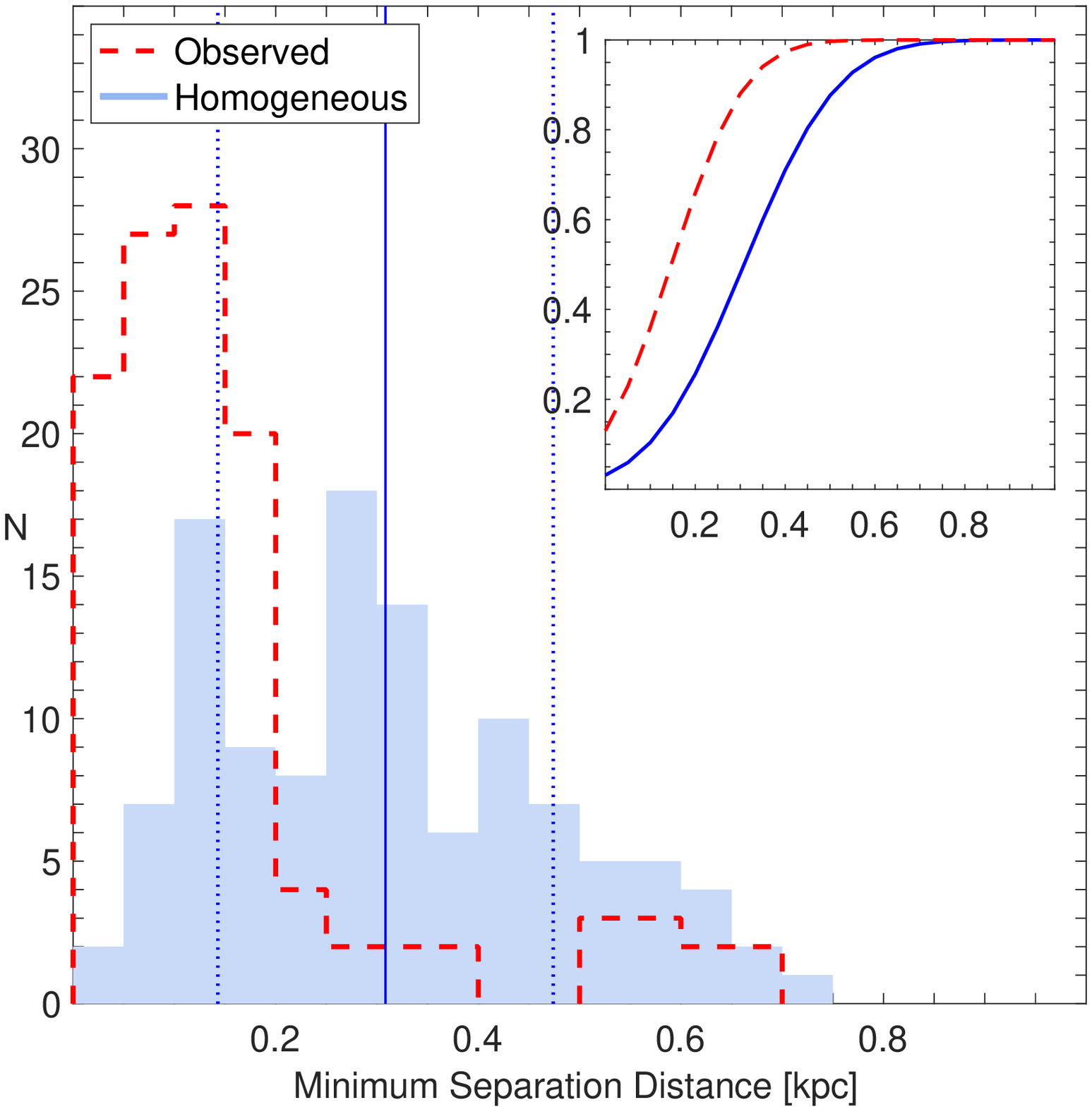}\hspace{10mm}\includegraphics[width=0.45\textwidth]{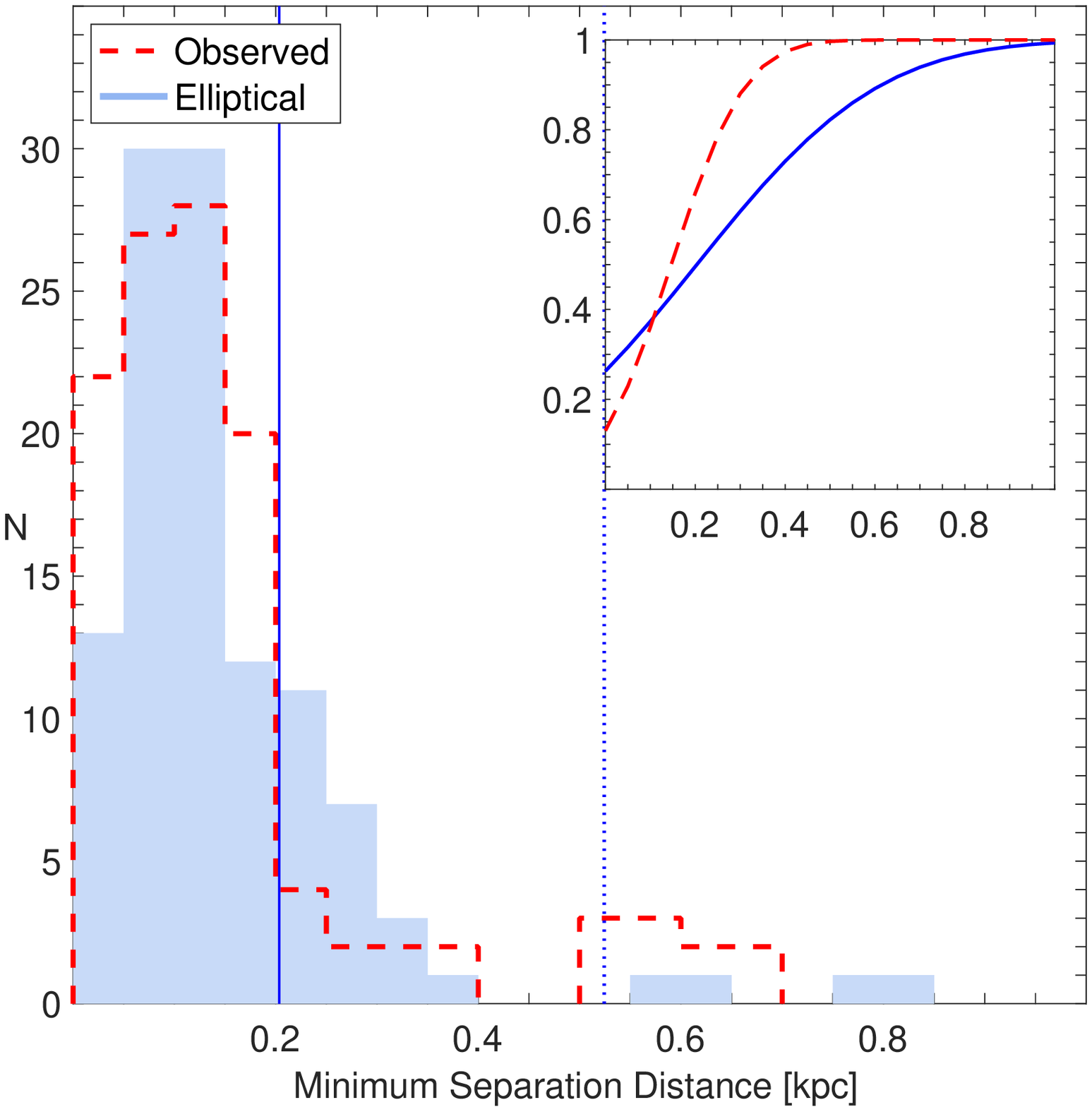}

\vspace{3mm}
\includegraphics[width=0.45\textwidth]{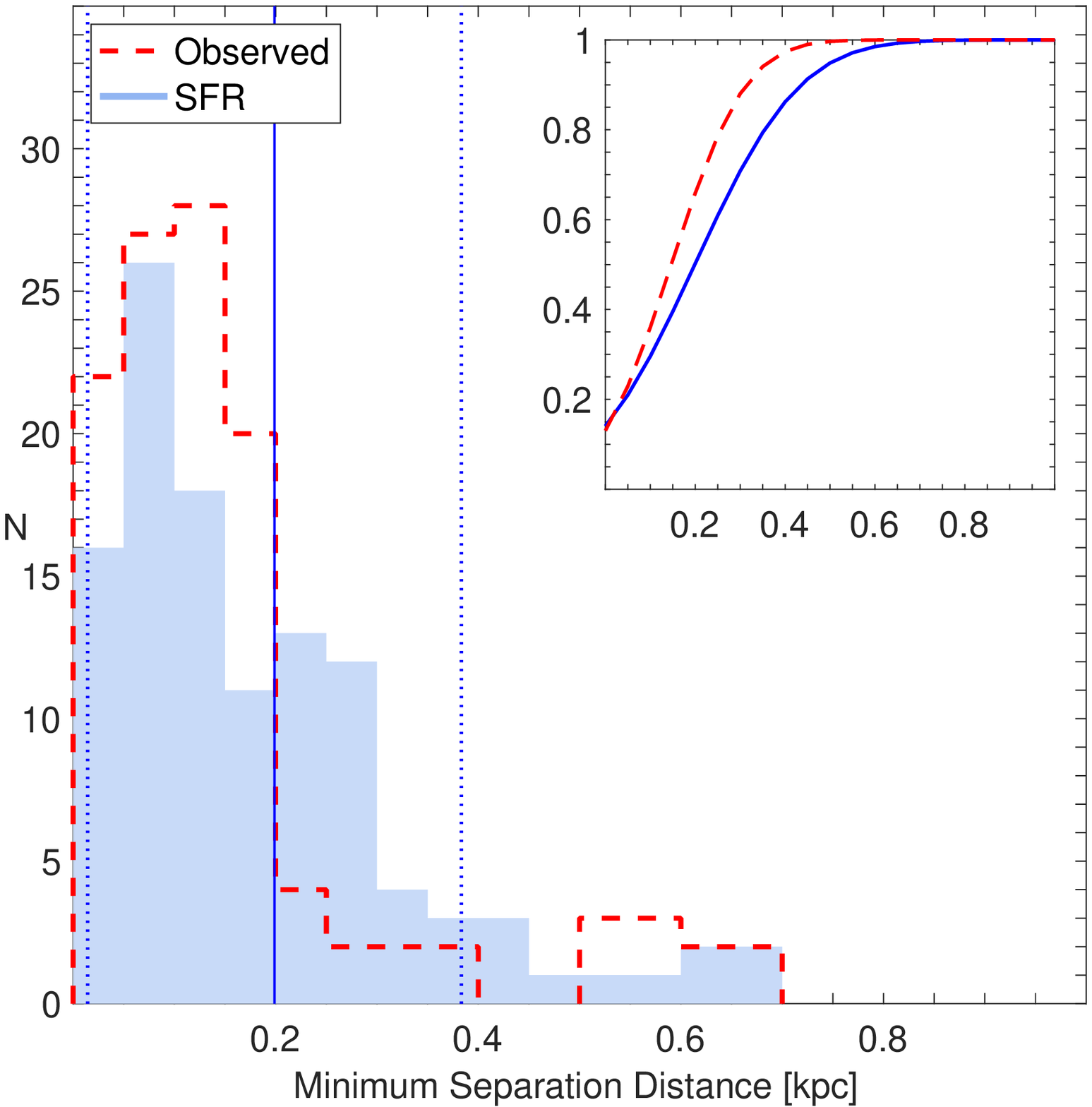}\hspace{10mm}\includegraphics[width=0.45\textwidth]{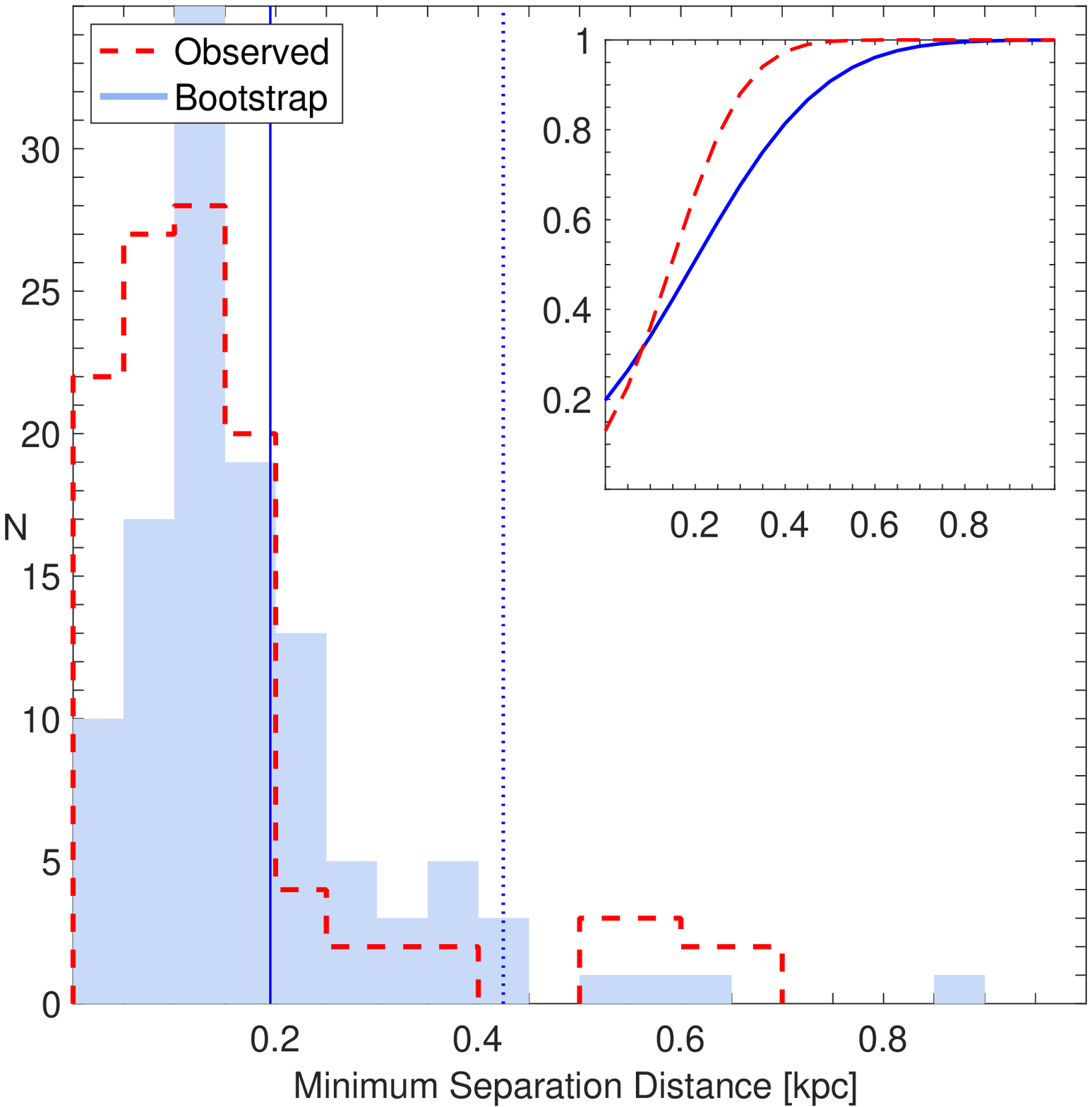}
\end{center}
\caption{Distribution of distances between an HMXB and its nearest OBA drawn from either the observed catalog (dashed red line) or from a randomized catalog (blue shaded region). Clockwise from the upper left, the panels show the four randomized OBA catalogs: Homogeneous, Elliptical, Bootstrap, and SFR. The vertical solid and dotted lines indicate the mean and standard deviation from a Gaussian fit to the distribution. The inset panels feature the same data over the same $x$-axis but plotted as cumulative distribution functions.}
\label{fig_sep_dist}
\end{figure*}
%-----------------------------Figure End--------------------------------

Four different mock catalogs were created each containing $10^5$ members whose sky coordinates were randomly generated according to some distribution function or resampling method. They are referred to as the ``randomized OBA catalogs:''

\begin{enumerate}

\item {\it Homogeneous}: the OBAs are placed randomly under the assumption of a uniform surface density. 

\item {\it Elliptical}: an elliptical function is fit to the observed OBA catalog and parameters (center and foci; lengths and angles of the major and minor axes) corresponding to the 90\%-containment region are extracted. Then, we randomly distribute OBAs across this ellipse according to a Gaussian profile: i.e., the number of objects peaks at the ellipse's center and gradually decreases outward. 

\item {\it SFR}:  the OBAs are randomly placed with a surface density weighted to the SFR of each region for the epoch corresponding to 40\,Myr ago \citep{rub18}, which \citet{ant10} cite as being the era from which the observed batch of HMXBs likely originates. We also used the SFR map of \citep{har04}, dividing the sky grid in the same way as these authors, then randomly placing OBAs according to the SFR for that epoch, and we found no significant differences in the results that are presented later.

\item {\it Bootstrap}: unlike the other randomized catalogs, where source coordinates were created based on an external constraint (e.g., mimicking the overall shape of the observed population), bootstrap sampling produces a randomized OBA catalog using only values found in the observed OBA catalog. New source coordinate pairs are generated by randomly assembling the right ascension value of one observed OBA with the declination value of another. The same coordinate value, and even the same coordinate pair, may appear multiple times in the final catalog, while others might not appear at all. Bootstrap resampling, which is frequently used in spatial correlation studies of extragalactic surveys \citep[e.g.][]{gil05,men09,kru10}, serves as a comparison to the methods above where we assume a specific morphology for the population.

\end{enumerate}

Examples of the four randomized OBA catalogs are shown in Figure\,\ref{fig_sky_random}. Figure\,\ref{fig_sep_dist} and Table\,\ref{tab_dist} compare the range of distances separating an HMXB and the nearest OBA belonging to the observed and randomized catalogs.

\subsection{The Spatial Correlation Function $\xi$}

Consider a member of Population\,1 located in a volume element $\delta V_{1}$. The probability $\delta P$ of finding a neighbor from Population\,2 in a volume element $\delta V_{2}$ separated by a distance $r$ is given by:

$$
\delta P = n_{1}n_{2}  \left [ 1 + \xi (r)  \right ] \delta V_{1}\delta V_{2}
$$

\noindent
where the number densities of each population are listed as $n_{1}$ and $n_{2}$, and where $\xi (r)$ is the spatial (or two-point) correlation function. If $\xi = 0$, then the equation yields a uniform probability, and so the spatial correlation function essentially describes the probability in excess of Poisson. \citet{pee80} provide the following estimator for $\xi$:

$$
\xi (r) = \frac{ D_{1}D_{2} }{ D_{1}R_{2} } -1
$$

The subscripts (1) and (2) will refer to HMXBs and OBAs, respectively. A member drawn from an observed catalog is designated $D$ for data while a member of a population drawn from a randomized catalog is labeled $R$. When two letters are combined, it signifies the normalized number of pairs of such type within a given volume element. In other words, $D_{1}D_{2} \equiv n^{2}N_{D_{1} D_{2}}(r)$ represents the number of pairs combining an observed HMXB with an observed OBA within a volume element of radius $r$, while $D_{1}R_{2} \equiv n N_{D_{1}R_{2}} (r)$ signifies the number of pairs combining an observed HMXB with an OBA from a randomized catalog. The ratio of random to observed data points is given by $n \equiv n_{R}/n_{D}$. 

Since we do not have reliable distance measurements for these objects, we assume that they are all located at the average SMC distance of 61\,kpc. This means that distances between members of the populations, and any relative motion, are only considered along the direction tangent to our line of sight. So instead of volume elements, we use surface elements for spatial binning. 

Around each HMXB, we draw 30 concentric annuli where each spatial bin has radial boundaries corresponding to $[ r - dr/2,\, r+ dr/2)$, and where $dr$ is the spatial bin size of $0.5^{\circ}$. This angular scale corresponds to a physical distance of around 500\,pc at the SMC distance. Different sizes were attempted for the spatial bins ($0.3^{\circ}$, $0.5^{\circ}$, and $1.0^{\circ}$), but we settled on $0.5^{\circ}$ since it was wide enough to permit enough counting statistics while also encompassing the average migration distance of 65\,pc that was found by \citet{coe05b} for HMXBs in the SMC. Within each spatial bin, we count the number of OBAs drawn from the observed catalog ($D_{1}D_{2}$) while keeping a tally of the members drawn from a randomized OBA catalog ($D_{1}R_{2}$). 

Counting in this manner for all HMXBs, we generate $\xi$ in each spatial bin out to 15\,kpc. To account for any unexpected spatial clustering or voids within the randomized catalog itself, we repeat the process $N = 10^{4}$ times (trials), where in each trial we select a new subset of OBAs from the randomized catalog. Each trial assumes an equal number of randomized OBAs as observed OBAs occupying equal sky areas, so $n=1$ in the estimator. We thus obtain a mean value of $\xi$ for each distance $r$ from a given HMXB.

Essentially, if a given spatial bin contains as many $D_{1}D_{2}$-pairs as $D_{1}R_{2}$-pairs, then $\xi = 0$ for that radius which means that the neighbor of a given HMXB is equally as likely to be drawn from the observed OBA catalog as it is from the randomized OBA catalog. On the other hand, if $\xi$ is significantly greater than 0, which will be referred to as a ``clustering signal,'' then OBAs near a given HMXB are more likely to be members of the observed (as opposed to the randomized) OBA catalog.

\begin{deluxetable}{ l l c c }
\tablewidth{0pt}
\tabletypesize{\scriptsize}
\tablecaption{Separation Distances (kpc) Between HMXBs and OBAs}
\tablehead{
\colhead{\noindent OBA Catalog} 	& 				& \colhead{Mean Distance} 	& \colhead{Minimum Distance\tablenotemark{a}} }
\startdata
Observed 						& Total			&	$3.62 \pm 0.97$		& 0.15 (0.13)				\\

	 						& Bar 			&	$2.56 \pm 0.49$		& 0.12 (0.09)				\\

	 	 					& Wing			&	$6.40 \pm 1.16$		& 0.46 (0.18)				\\

\hline\\
Randomized					& Homogeneous	&	$5.40 \pm 0.64$		& 0.31 (0.17)				\\

							& Elliptical 		&	$3.34 \pm 1.20$		& 0.20 (0.32)				\\

							& SFR			&	$3.49 \pm 1.30$		& 0.20 (0.18)				\\

							& Bootstrap		&	$3.34 \pm 1.05$		& 0.19 (0.23)				\\

\enddata
\tablenotetext{a}{This corresponds to the mean and standard deviation from a Gaussian fit to the histograms in Figure \ref{fig_sep_dist}.}
\label{tab_dist}
\end{deluxetable}

\citet{lan93} propose an alternate estimator of the spatial correlation function:

$$
\xi (r) = \left( D_{1}D_{2} - D_{1}R_{2} - D_{2}R_{1} + R_{1}R_{2} \right) \frac{1}{{R_{1}R_{2}}}
$$

In this function, the number counts within each spatial bin are also considered for HMXBs that are randomly distributed in the same manner as the OBAs. These terms appear as $D_{2}R_{1} \equiv nN_{D_{2}R_{1}}(r)$ and $R_{1}R_{2} \equiv N_{R_{1}R_{2}}(r)$. This accounts for any effects caused by the choice of survey boundaries or incompleteness in the observed populations. After ensuring the consistency of the results for both the \citet{pee80} and \citet{lan93} estimators of $\xi$, we elect to use the latter whose variance is expected to be closer to Poissonian.

\subsection{Error Analysis}

Studies of the correlation function applied to cosmological surveys show that the uncertainty associated with one spatial bin of the survey is interdependent on the other bins \citep[e.g.,][and references therein]{lin86,ber94,nor09,men09}. Furthermore, these authors show that the choice of randomization method will affect the uncertainty on $\xi$; whether internal methods are employed (e.g., the Bootstrap catalog), or whether external estimates are used (e.g., the Elliptical and SFR catalogs). 

Errors calculated with internal methods can account for systematic biases, e.g., detection limits of the telescope, and observation strategies. Such biases are not accounted for in external methods where we make assumptions about the underlying physical processes that created the distribution. On the other hand, internal estimates can be hindered by the limited size of the original dataset. Variances from bootstrapping can be larger by a factor of 2 or more compared with those of Poisson statistics, but they appropriately describe the correlation function at small distance scales, especially when oversampling \citep[e.g.,][]{nor09}. Since our data are interdependent, including across spatial bins, we employ moving-block bootstrapping where overlapping blocks of varying length are resampled to create the Bootstrap catalog \citep{kun89,loh08}.

Therefore, additional tests are performed in order to better ascertain the uncertainty on $\xi$. For each randomization method, we determine the average $\xi$ for each of the 30 spatial bins over all $10^{4}$ trials. Then, we generate the $30 \times 30$ covariance matrix $C$ whose elements are:

$$
C_{ij} = \frac{1}{N-1} \sum_{k=1}^{N}{(\xi_{i}^{k} - \bar{\xi}_{i}) (\xi_{j}^{k} - \bar{\xi}_{j}) }
$$

\noindent
where $\bar{\xi}$ is the mean:

$$
\bar{\xi}_{i} = \frac{1}{N}\sum_{k=1}^{N}{ \xi_{i}^{k} }
$$

\begin{deluxetable}{ l c c c c }
\tablewidth{0pt}
\tabletypesize{\scriptsize}
\tablecaption{First Five Eigenvalues of the Correlation Matrix of $\xi$.}
\tablehead{
\colhead{ \,		} 	& \colhead{Homogeneous} 	& \colhead{Elliptical}  & \colhead{SFR}	& \colhead{Bootstrap} 			}
\startdata

$\lambda_{1}$ 			&	17.1570				& 21.1065			& 21.9915			& 18.7840			 \\

$\lambda_{2}$ 			&	6.6233				& 4.4770			& 4.1303			& 5.7872			 \\

$\lambda_{3}$ 			&	2.4948				& 1.4395			& 1.4877			& 1.5210			 \\

$\lambda_{4}$ 			&	1.3956				& 0.7997			& 0.4570			& 1.0844			 \\

$\lambda_{5}$ 			&	0.5331				& 0.3335			& 0.2356			& 0.4458	 		\\

\enddata
\label{tab_eigen}
\end{deluxetable}

\noindent
and where $\xi_{i}^{k}$ denotes the value of $\xi$ in spatial bin $i$ for trial $k$. Standard deviations on $\xi$ are collected directly from the square root of the diagonal elements of the covariance matrix. 

The correlation matrix $R$ permits a direct check of the dependence of each of the 30 spatial bins on the other. Each element of the correlation matrix can be constructed according to:

$$
R_{ij} = \frac{C_{ij}}{\sqrt{C_{ii}C_{jj}}}
$$

The elements of $R$, when given in terms of their absolute values, decrease gradually from 1 (full correlation) to 0 (no correlation) when moving away from the diagonal along the same row or column. As discussed in the next section, there remains significant bin-to-bin correlation off diagonal at distance scales of 5--15\,kpc. We are interested primarily with clustering at small distance scales (i.e., within about 1\,kpc), so this will not affect our conclusions. 

Eigenvalues are gathered from each correlation matrix; and the first five components (of 30) are listed in Table\,\ref{tab_eigen}. While the sum of the eigenvalues equals the dimension as expected, it is clear that the primary eigenvalue, and to a lesser extent the next 4 components, dominate the correlation matrix made with each randomization method. 

As a{n additional} test, we introduce perturbations to the data by moving each HMXB in either a random direction or towards the nearest OBA, and recalculate $\xi$ in the normal manner. We  show later that $\xi$ remains stable to small-scale perturbations to the data on the order of a few hundreds of parsecs of physical distance.

%-----------------------------Figure Start------------------------------
\begin{figure}[!t]
\begin{center}
\vspace{0mm}
\includegraphics[width=0.45\textwidth]{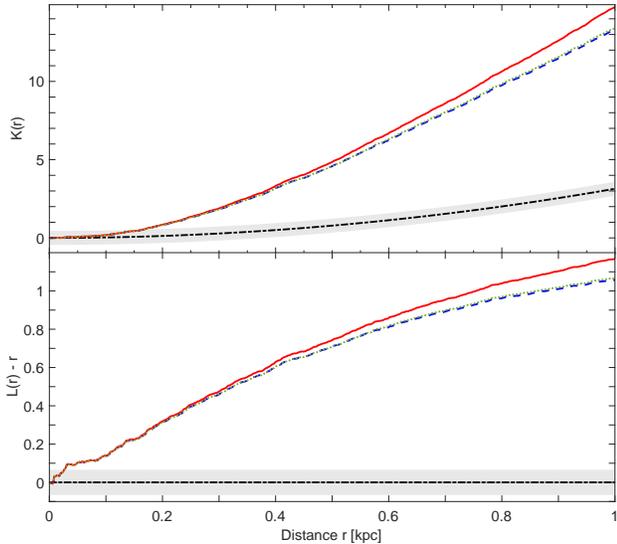}
\end{center}
\caption{Ripley's $K$ function (upper panel) and its associated $L$ function (lower panel) are illustrated for cross-pair distances between observed HMXBs and OBAs, with isotropic (dashed blue curve), translational (solid red curve), and border (dotted green curve) corrections applied. The dot-dashed curve refers to the Poisson expectation with the shaded region indicating the range of $10^{3}$ simulations. }
\label{fig_KLfunction}
\end{figure}
%-----------------------------Figure End--------------------------------

Finally, we applied Ripley's $K$ (cross-)function to the observed HMXB and OBA catalogs \citep[e.g.,][]{fei12}:

$$
K_{12} (r) = \left(n_{1} n_{2} A \right)^{-1} \sum_{i} \sum_{j} w(1_{i},2_{j})\,I(d_{1_{i},2_{j}} < r)
$$ 

As before, the labels 1 and 2 refer to HMXBs and OBAs, respectively, with their number densities given by $n_{1}$ and $n_{2}$ in the survey area $A$. The term $d_{1_{i},2_{j}}$ is the distance between the $i$th HMXB and the $j$th OBA. The indicator function $I$ has a value of 1 if this distance is within a circle of radius $r$ and 0 otherwise. To include edge corrections, the term $w(1_{i},2_{j})$ represents the fraction of this circle's circumference that resides within $A$. If the HMXB and OBA populations are spatially independent, then $K_{12} (r) = \pi r^{2}$. An alternate form is the $L$ (cross-)function:

$$
L_{12} (r) = \sqrt{\frac{K_{12} (r) }{\pi}}
$$

When the distance $r$ is subtracted from it, the $L$ function conveniently reduces to 0 under spatial independence. Both functions show significant departure from the values expected for complete spatial randomness (Fig.\,\ref{fig_KLfunction}). This strongly suggests spatial clustering between the HMXB and OBA populations for pair counts in circular areas whose radii are greater than about 0.1--0.2\,kpc.

Unless specified otherwise, results in this work are cited at 90\% confidence.

%-----------------------------Figure Start------------------------------
\begin{figure*}[!t]
\begin{center}
\includegraphics[width=0.45\textwidth]{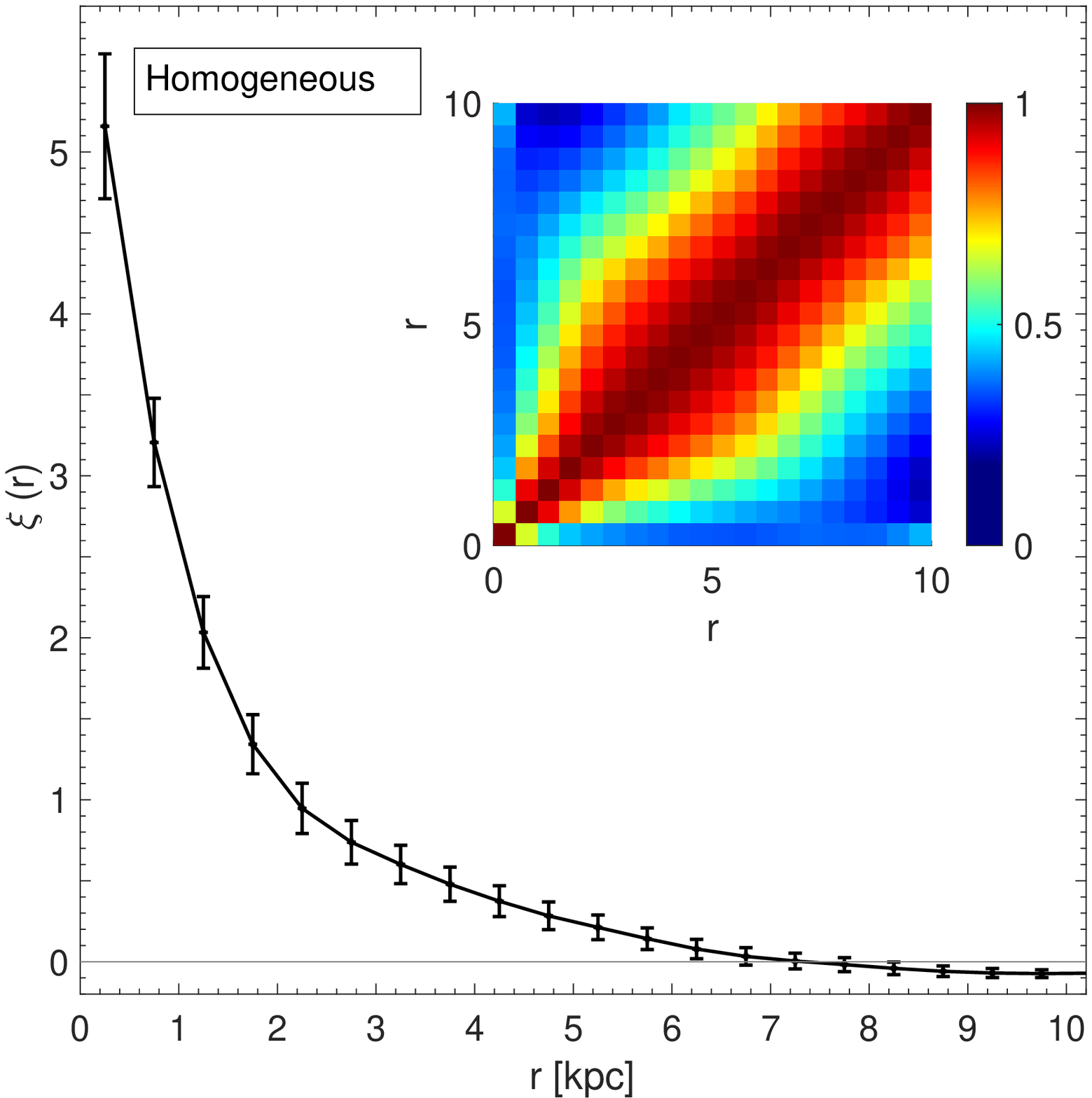}\hspace{10mm}\includegraphics[width=0.45\textwidth]{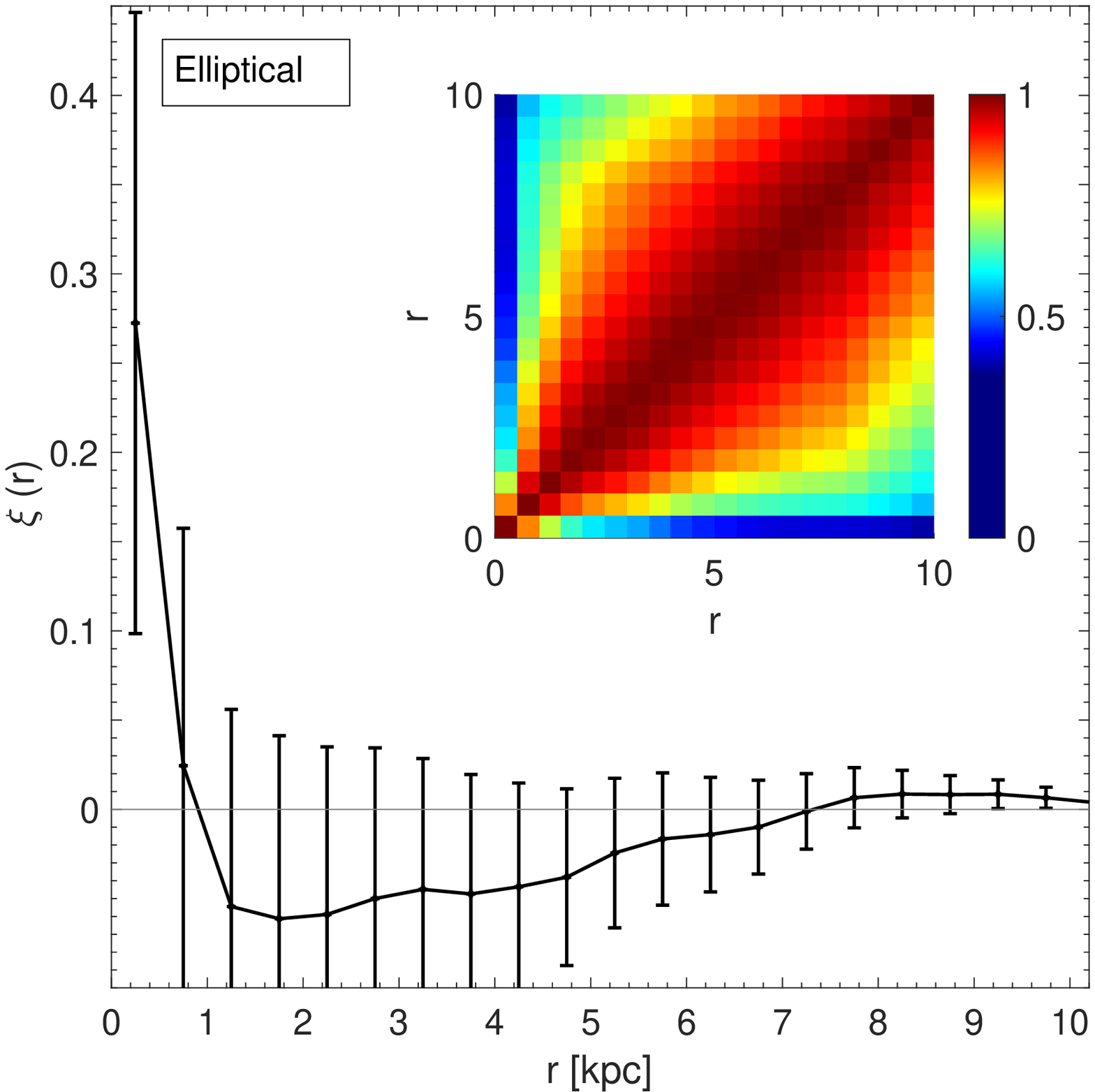}

\vspace{3mm}
\includegraphics[width=0.45\textwidth]{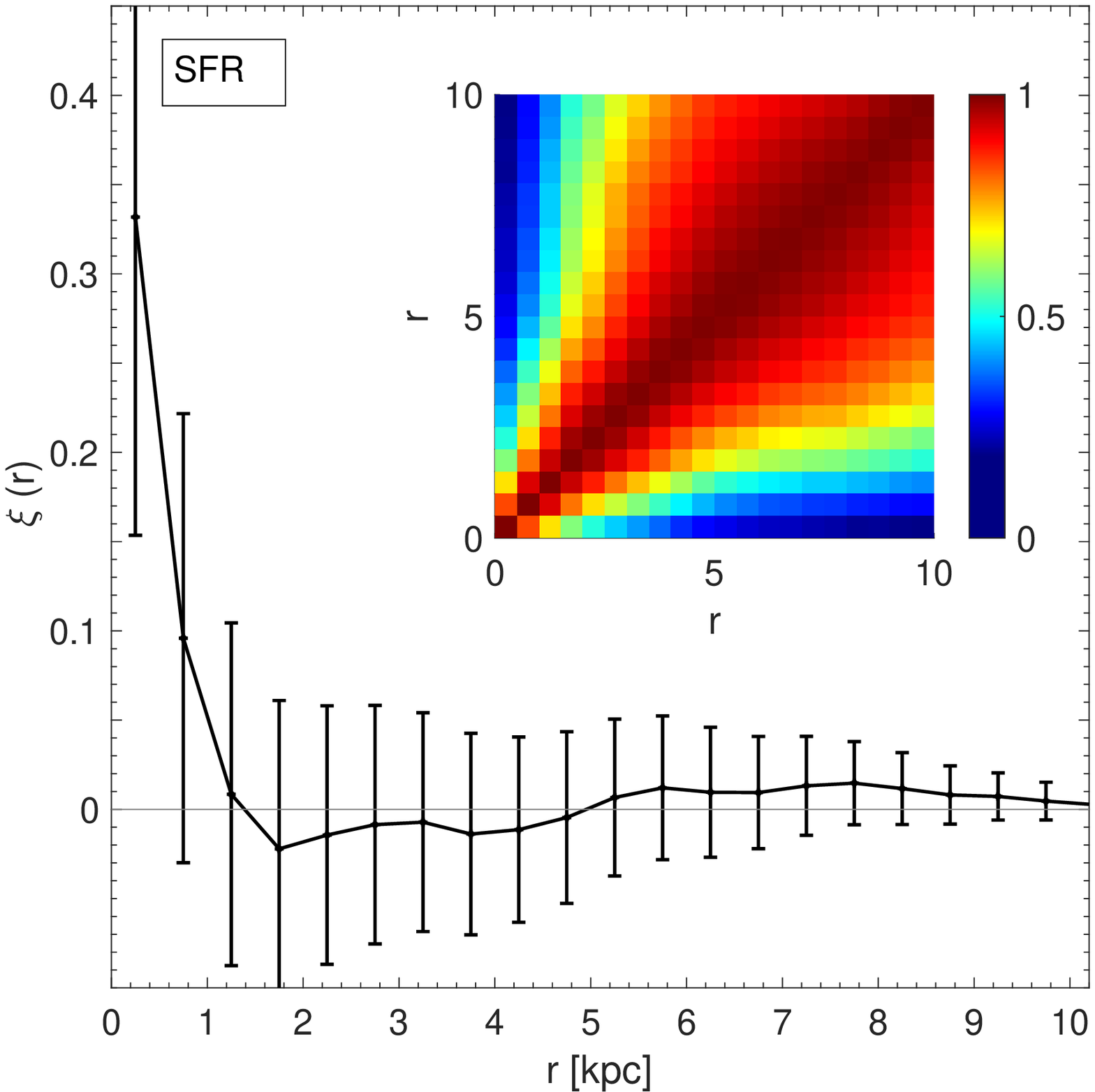}\hspace{10mm}\includegraphics[width=0.45\textwidth]{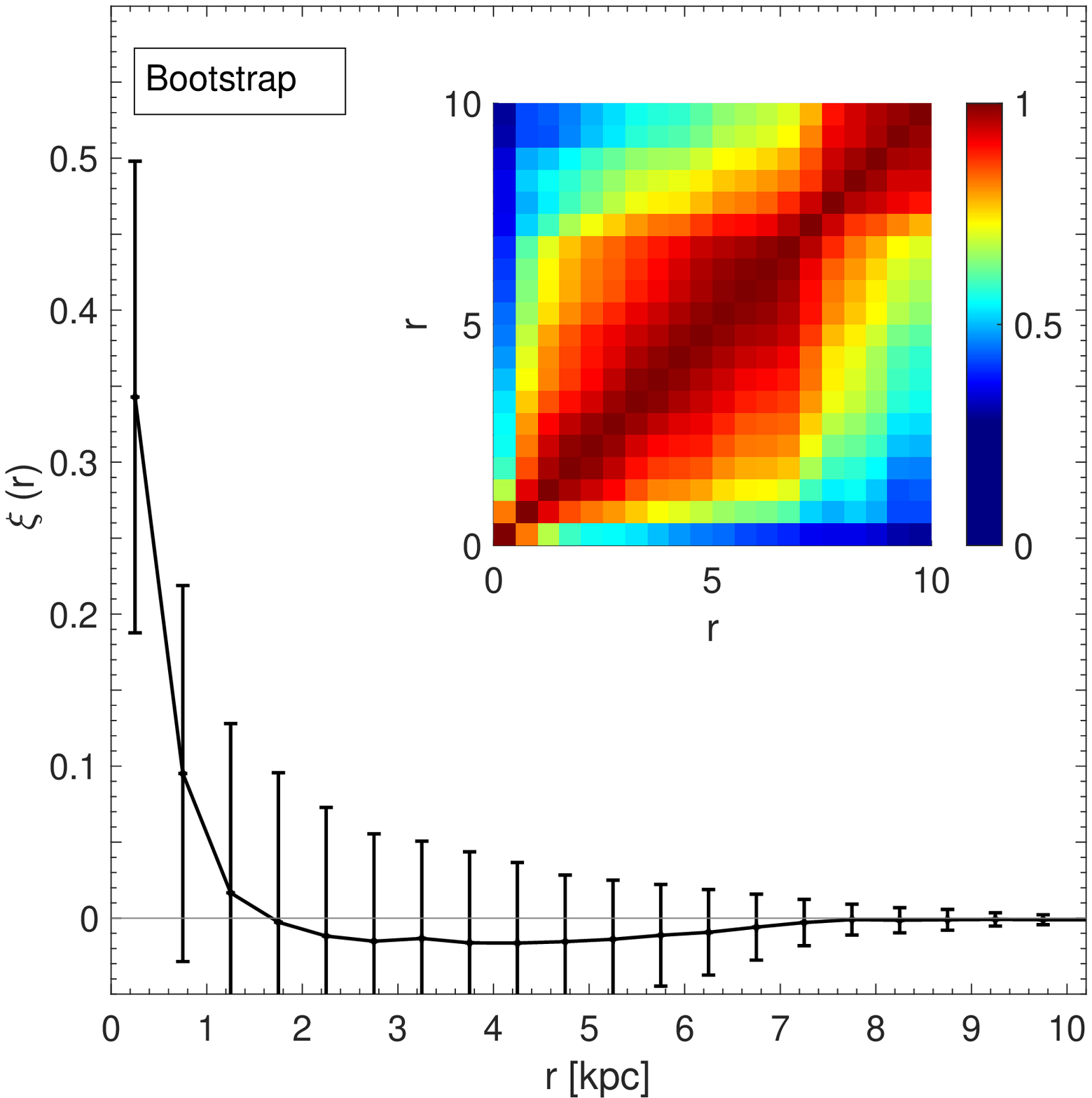}
\end{center}
\caption{Spatial correlation functions $\xi$ derived by comparing HMXBs to the observed OBA population and to one of four randomized OBA catalogs, where clockwise from the upper left, the panels show distributions corresponding to Homogeneous, Elliptical, Bootstrap, and SFR. The inset in each panel shows the correlation matrix for $\xi$ whose eigenvalues are listed in Table\,\ref{tab_eigen}.}
\label{fig_xi}
\end{figure*}
%-----------------------------Figure End--------------------------------

\section{Results \& Discussion}
\label{sec_res}

\subsection{Spatial Clustering}

Figure \ref{fig_xi} presents the spatial correlation function $\xi$ as a function of the distance $r$ from a given HMXB, for each of the four randomized OBA catalogs. In all four cases, $\xi$ was consistent with 0 at large distances from a given HMXB ($r \gtrsim 5$\,kpc). This is expected since sky regions far from a given HMXB should contain equivalent numbers of OBAs drawn from the observed catalog as from the randomized catalog.

As one approaches a given HMXB ($r \lesssim 5$\,kpc), $\xi$ deviates significantly from 0 when the randomized OBA catalog follows a homogeneous distribution (upper left panel of Fig.\,\ref{fig_xi}). The spatial correlation function is significantly greater than 0 for distances up to 3\,kpc, with a peak of $15\sigma$ for distances less than 0.5\,kpc from a given HMXB. The Homogeneous catalog has a distribution that is Poissonian by design, so this means there is an overabundance of observed OBAs in close proximity to an HMXB. 

In other words, the observed OBA catalog offers a better match to the HMXBs than would be expected from a random distribution. This is evidence of spatial clustering which reinforces the link between OBAs and HMXBs. This clustering was previously established for smaller samples in the SMC by, e.g., \citet{coe05b} and \citet{ant10}, based on the minimum separations between individual members of the observed populations as shown in Fig.\,\ref{fig_sep_dist} and Table\,\ref{tab_dist}. However, this is the first time that the clustering is confirmed with the spatial correlation function and it does not require us to assume that a given HMXB is linked with the nearest OBA. 

Randomizing the OBA catalog to follow an Elliptical distribution yields a $\xi$ that remains consistent with 0 for all distances from a given HMXB, with an insignificant ($2\sigma$) uptick at the shortest distances (upper right panel of Fig.\,\ref{fig_xi}). Since the spatial correlation function does not find significant clustering at any distance scale, this suggests that the Elliptical OBA catalog is a good approximation of the observed distribution of OBAs. This is not surprising given that this model's parameters were based on a fit to the observed OBA catalog. 

When the randomization profile is based on mimicking the SFR of the SMC, the $\xi$ that results has values that are consistent with 0 for all distances from a given HMXB, with a low significance ($3\sigma$) deviation from 0 in the first spatial bin (lower left panel of Fig.\,\ref{fig_xi}). As with the Elliptical catalog, the overall shape of the SFR catalog appears to be a suitable proxy for that of the observed OBA catalog. These results imply that the actual locations of the randomized OBAs in the SFR catalog do not matter as much as does the overall shape of the distribution. Members of the SFR catalog could change their coordinates, which they do change multiple times over all the trials, but as long they are placed according to a number density that is proportional to the star-formation rate, then the spatial correlation function can not tell the difference between the SFR catalog and the observed OBA catalog. Taken together, this suggests that places where massive stars are forming today, i.e., what we call the observed OBA catalog, are essentially the same as the regions of the SMC that featured a peak outburst of star formation 40\,Myr ago \citep{ant10,ant16}.

The Bootstrap catalog generates a $\xi$ function with a small, but significant ($4\sigma$), clustering signal in the first spatial bin (lower right panel of Fig.\,\ref{fig_xi}). The clustering signal remains above $3\sigma$ in the first bin even if the variance is increased by 50\%. %Of the randomized OBA catalogs, the Bootstrap catalog has a distance distribution that most closely resembles that of the observed OBA catalog (Table\,\ref{tab_dist} and the cumulative distribution function in Fig.\,\ref{fig_sep_dist}), and its correlation matrix has a larger principal component implying a tighter correlation between the value of $\xi$ and $r$ (Table\,\ref{tab_eigen}). 

In summary, there is a strong spatial link between observed HMXBs and observed OBAs as attested by the $\xi$ made with the Homogeneous catalog, and it is of low significance with the Bootstrap catalog, and possibly with the SFR Catalog. The spatial correlation functions made with the Elliptical and SFR catalogs do not show any significant clustering which is not surprising given that these catalogs are based on fitting the observed OBA distribution in some way. 

Clustering between the observed HMXB and OBA populations takes on three visible forms: it can be pair-wise whereby an HMXB is found within a few tens or, at most, a few hundred parsecs from a neighboring OBA (this is the case for a subset of the HMXBs); it can be group-wise whereby a small group of HMXBs is found within a few kpc of a group of OBAs; and it can be global meaning that the overall morphologies of the populations are compatible. Using the Homogeneous catalog, the spatial correlation function shows evidence of pair-wise and group-wise clustering with the signal remaining significant to 2--3\,kpc. With the Elliptical and SFR catalogs, there is global compatibility, and even some group-wise clustering, but there does not appear to be enough pair-wise clustering to produce a significant signal. If the clustering signal found with the Bootstrap catalog is real, then it could suggest some pair-wise clustering in addition to some group-wise clustering and global similarity. 

%-----------------------------Figure Start------------------------------
\begin{figure}[!t]
\begin{center}
\vspace{0mm}
\includegraphics[width=0.45\textwidth]{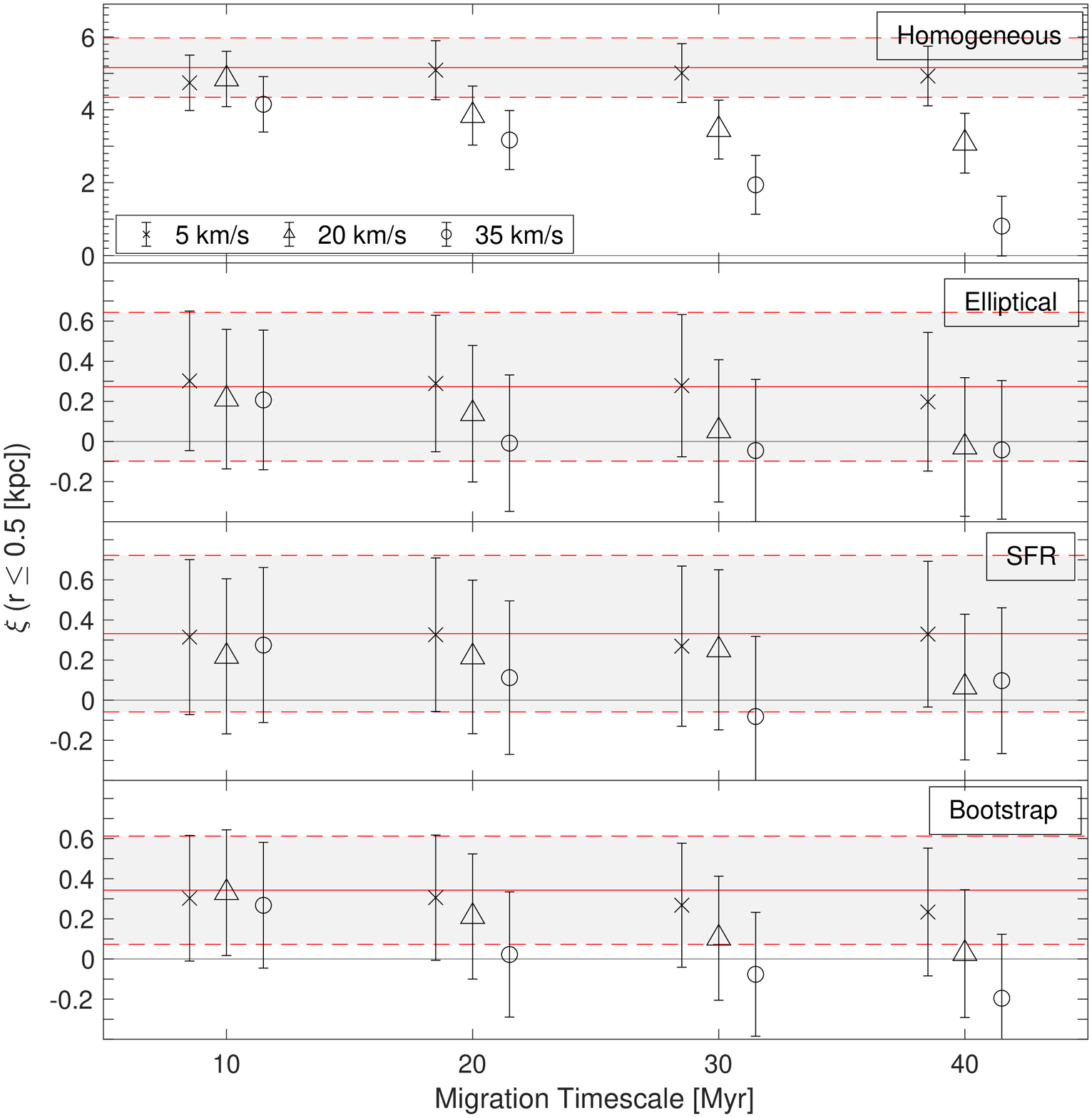}
\end{center}
\caption{Values for the spatial correlation function $\xi$ (with 3$\sigma$ error bars) as a function of the migration timescale for small distances from an HMXB ($r \le 0.5$\,kpc). Each HMXB was moved in a random direction at some velocity (crosses, triangles, and circles denote 5, 20, and 35\,km\,s$^{-1}$, respectively) during timescales of 10, 20, 30, and 40 Myr. The original, unperturbed data ($v=0$\,km\,s$^{-1}$) is shown as a solid red line while the dashed red lines represent 3$\sigma$ confidence levels (shaded in gray to denote the boundaries). }
\label{fig_xi_shift_rand}
\end{figure}
%-----------------------------Figure End--------------------------------

%-----------------------------Figure Start------------------------------
\begin{figure}[!h]
\begin{center}
\vspace{0mm}
\includegraphics[width=0.45\textwidth]{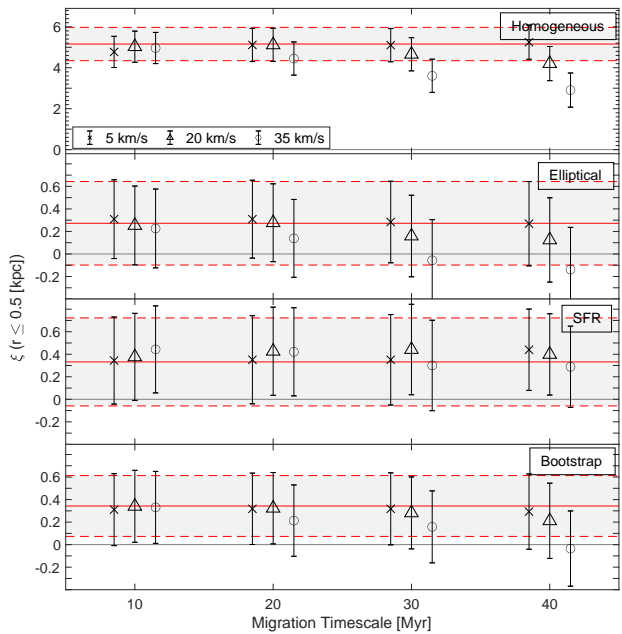}
\end{center}
\caption{Same as Fig.\,\ref{fig_xi_shift_rand}, but in this case, each HMXB is moved in the direction of its nearest OBA. }
\label{fig_xi_shift_near}
\end{figure}
%-----------------------------Figure End--------------------------------

To test the degree to which the pair-wise and group-wise clustering influence the correlation, each HMXB was displaced in a random direction assuming some velocity (5, 20, and 35 km\,s$^{-1}$) and some timescale (10, 20, 30 and 40\,Myr). Averaging $\xi$ over all trials leads to the results shown in Fig.\,\ref{fig_xi_shift_rand}. Only $\xi$ from the closest region around a given HMXB is shown ($r\le0.5$\,kpc). The figure illustrates the stability of the spatial correlation function's values (indicated by the horizontal dashed lines showing the 3$\sigma$-confidence boundaries) to small perturbations of the data (represented by the symbols). In nearly all cases, the perturbed and unperturbed $\xi$ values were statistically consistent with each other. The only exceptions were for high velocities ($v=20$ or $35$\,km\,s$^{-1}$) and long timescales ($t=30$ and $40$\,Myr) where the significance of the clustering signal decreases for the Homogeneous catalog. Essentially, these perturbations broaden the HMXB distribution on the sky. This enhanced scatter makes the HMXB distribution more similar to the Homogeneous catalog, which leads to a decrease in the value of $\xi$. 

In an alternate test, we moved each HMXB towards its nearest OBA under the same assumptions of velocity and timescale as before, and generated $\xi$ in the usual way over multiple trials. By moving the observed populations closer together, the clustering significance is expected to increase if pair-wise clustering dominates over group-wise or global clustering. However, the clustering signal did not change in a significant way for any randomized catalog (Fig.\,\ref{fig_xi_shift_near}). Taken with the result of the previous test, where the HMXBs were moved around randomly, this suggests that the spatial correlation function gets its significance mostly from group-wise and global clustering, with pair-wise clustering being a less important factor. 

Structurally, the SMC can be divided into two regions: the Bar which hosts the majority of the dwarf galaxy's massive stars and HMXBs; and the Wing which roughly corresponds to the sky area in Fig.\,\ref{fig_sky} having R.A. $\gtrsim 18^{\circ}$ and Decl. $\lesssim -72.5^{\circ}$ \citep[see also][]{rub18}. \citet{oey18} showed that objects in these regions can be distinguished kinematically with the Wing moving away from the Bar and towards the LMC at $\sim 64$\,km\,s$^{-1}$. The clustering signal remains at a significance of 16$\sigma$ and 4$\sigma$ for the Homogeneous and Bootstrap catalogs, respectively, when the correlation function is limited to the 106 HMXBs in the Bar. The correlation function applied only to the 9 HMXBs in the Wing generates error bars that are too large to draw meaningful conclusions. In the Bar, the average minimum distance between an HMXB and an OBA is 120$\pm$90\,pc which is a factor 2--4 less than in the Wing (450$\pm$180\,pc).

\subsection{Constraints on Kick Velocities}

The main reason to study the spatial clustering of HMXB and OBA populations is to determine the magnitude of the natal kick of the HMXB as it moves away from its birthplace. Migration of this magnitude is not observable on human timescales, so we do not know which OBA is the birthplace of each HMXB, and it is not necessarily the closest one on the sky. However, knowing how clustered the two populations are can tell us the typical migration distance. if we have an idea of how old these objects are, we can use this distance to determine a kick velocity. Gravitational ejection, cluster outflows, recoil due to anisotropic mass transfer, and asymmetric supernovae can impart velocity to a young HMXB displacing it from its parent cluster \citep{bla61,pov67,shk70,pfl10}.

Based on a Gaussian fit to the distance distributions (Fig.\,\ref{fig_sep_dist} and Table\,\ref{tab_dist}), we find that the distance separating an HMXB from the nearest observed OBA is 150\,pc, on average, with a 90\%-confidence upper limit of 300\,pc. These minimum distances are even smaller in the Bar (120$\pm$90\,pc). Twenty percent of the HMXBs in the sample have a minimum distance below 50\,pc, with the smallest values on the order of 10\,pc. This sets a lower limit on the migration distance. These values are consistent with that of \citet{coe05b} who found 65\,pc from a sample of 17 HMXBs. 

Among these minimum distances, the highest values are in the range of 500--700\,pc. We can compare this with the value given by the spatial correlation function which is designated by the largest radius for which $\xi$ remains significantly greater than 0. Using the Homogeneous catalog, the clustering signal is significant out to a distance of 2--3\,kpc from a given HMXB which is equivalent to $\sim$2$^{\circ}$--3$^{\circ}$ on the sky. This represents an unrealistically large value for a typical migration distance. When using the Bootstrap catalog, the clustering signal is significant in the first bin only which suggests a maximum migration distance of $\sim$500\,pc, consistent with the value from the distance distribution. This sets an upper limit on the migration distance, and so the range of distances separating an HMXB from a neighboring OBA is $\sim$ 10--700\,pc with a mean of 150\,pc. 

Around 40\,Myr ago, the SMC underwent a particularly intense episode of star formation which produced dozens of massive stellar binaries per Myr, and per arcmin squared \citep{dra06,ant10}. Depending on the initial masses of the stellar binary, another 5--20\,Myr are needed for the system to evolve into an HMXB that kicks itself out of its OBA \citep{sav77,bel08}. This sets theoretical lower and upper bounds on the migration timescale of 20--35\,Myr. 

These timescales lead to kick velocities of 2--34\,km\,s$^{-1}$ over the range of minimum distances found here. Since this is purely the transverse velocity component, the true velocity could be twice as large, on average. This agrees with estimates of the kick velocity in HMXBs from the SMC found by \citet{coe05b,ant09,ant10,oey18} with smaller samples, and is consistent with the prediction of $15\pm6$\,km\,s$^{-1}$ for BEXBs given by \citet{van00}. It also agrees with the mean transverse velocity of $11.3\pm6.7$\,km\,s$^{-1}$ that was found for 13 Galactic BEXBs in the \hipp\ catalog \citep{che98}. Binary systems that survive the SN phase to later remain bound as an HMXB tend to have low velocities that are attributable to natal kicks, whereas large migration velocities are acquired from the gravitational potential energy of the cluster. The average velocity that we find is lower than expected from cluster ejection where simulations show velocities over 80\,km\,s$^{-1}$ with rare cases reaching several hundred\,km\,s$^{-1}$ \citep{per12}.

More than half the HMXBs in the SMC have a kick velocity that is significantly less than that of their counterparts in the Milky Way, where the average velocity was found to be 100$\pm$50\,km\,s$^{-1}$ at 90\% confidence \citep{bod12}. Accepting at face value the maximum migration distance of 2--3\,kpc from the clustering signal of the Homogeneous catalog, this returns kick velocities of 50--150\,km\,s$^{-1}$ that are in line with velocities of HMXBs in the Milky Way. Yet these would be the most extreme cases, and are not representative of the majority.

It is possible to derive kick values that are more consistent with those of the Milky Way if one considers the average HMXB lifetime to be much lower (e.g., at 10 Myr). A factor of 4 reduction in average lifetime corresponds to a factor of 4 increase in the kick velocity, and so the average kick velocity of HMXBs in the SMC will be consistent (within the errors) with those of the Milky Way only if the HMXBs are $\lesssim$10 Myr old. However, this would be too short a timescale for stellar evolution models and it is not consistent with the expectation that the majority of these systems are byproducts of that starburst episode from 40\,Myr ago.

Thus, we find that HMXBs in the SMC are migrating less quickly than those of the Milky Way. In our Galaxy, the known HMXB population is nearly evenly split between the classes of BEXBs and SGXBs \citep{wal15,kre19}. In the SMC, the HMXB population is almost exclusively composed of BEXBs (113/\numHMXB). \citet{van00} showed that BEXBs have kick velocities that are around 3 times lower than those of SGXBs, though there was no notable difference in how these classes were clustered with OBAs in the Milky Way \citep{bod12}. Using this velocity factor under a hypothetical scenario in which the SMC population is evenly split between BEXBs and SGXBs, then only the fastest systems in the SMC (70\,km\,s$^{-1}$) would overlap with the lower bound of the Milky Way population. So some other effect must be responsible for constraining the average velocity of HMXBs in the SMC.

It is known that the lower average metallicity of stars in the SMC led to a preponderance of BEXBs over SGXBs, particularly those with low kick velocities \citep[e.g.,][]{lin09,lin10,ant10,kaa11,bas16}. At lower metallicity, a higher fraction of electron-capture SNe is expected, and this disproportionally generates BEXBs having low values for the kick velocity ($\lesssim 50$\,km\,s$^{-1}$). This effect, cited by \citet{ant16} to explain the lower velocities in the SMC compared with the LMC, is still present with 43 additional HMXBs in this work. In a recent study \citep{oey18}, 15 SMC HMXBs showed relatively little dispersion in their mean velocity of $17$\,km\,s$^{-1}$ compared with non-compact binary systems which suggests the former are the product of less energetic supernova kicks.

An alternative explanation for the low average kick value is that some of the HMXBs in the SMC harbor black holes whose formation via direct collapse failed to produce a natal kick \citep{bel20}. As of this study, no black hole has ever been confirmed among the HMXBs in the SMC. Assuming that the non-pulsating HMXBs in the SMC all host black holes, they would constitute a minority of the population, so this process has a small or negligible impact on the average kick that we measure for the whole population.

\section{Summary \& Conclusions}
\label{sec_conc}

In this study, we generated the first two-point cross-correlation function between HMXBs and OBAs belonging to the SMC. In order to assess the spatial distributions of these related populations, we built comparison catalogs in which the systems were distributed in a random fashion. Four catalogs with randomized distributions were considered: homogeneous; a Gaussian ellipsoid; a distribution that mimics the recent star-formation history of this galaxy; and a bootstrap resampling. 

The spatial distributions of the HMXB and OBA populations were found to be significantly correlated when compared with a randomized catalog in which the systems are spread homogeneously across the field (15$\sigma$), or when following bootstrap resampling (4$\sigma$). This reinforces the link expected between the observed HMXB and OBA populations. On the other hand, no correlation was seen when the randomized catalogs assume the Elliptical and SFR distributions. This is not surprising since these catalogs were made by fitting some characteristic of the observed population. 

Our results indicate that in the SMC, an HMXB inherits an average natal kick velocity of 2--34\,km\,s$^{-1}$, which is consistent with values cited in the literature. This is considerably less than the average kick velocity of HMXBs in the LMC and Milky Way. One factor driving this discrepancy is that BEXBs, which inherit smaller average kick velocities than do SGXBs, are far more common among the HMXBs in the SMC. This is due to the lower average metallicity in the SMC which favors the creation of BEXBs over SGXBs. This lower average metallicity also contributes to the discrepancy by leading to a higher probability of electron-capture supernovae which produce less energetic kicks. Thus, the galactic environment plays a role in setting the natal kick which is then perceived as smaller migration velocities in the observed HMXB population.

\acknowledgments
The authors are grateful to the anonymous Scientific Referee and to the Statistics Referee for their comments and suggestions which improved the quality of the manuscript. AB, RA, and BJ acknowledge partial funding of this study by Georgia College's Mentored Undergraduate Research And Creative Endeavors (MURACE) Grant. VA acknowledges financial support from NASA/Chandra grants GO3-14051X, AR4-15003X, NNX15AR30G and NASA/ADAP grant NNX10AH47G. JR acknowledges partial funding from the French Space Agency (CNES) and the French National Program for High-Energy Astrophysics (PNHE). This research has made use of: data obtained from the High Energy Astrophysics Science Archive Research Center (HEASARC) provided by NASA's Goddard Space Flight Center; NASA's Astrophysics Data System Bibliographic Services; the SIMBAD database operated at CDS, Strasbourg, France; Matlab's Statistics \& Machine Learning Package; and R's ``spatstat'' Package available at \texttt{\small{https://CRAN.R-project.org/package=spatstat}}.

\bibliographystyle{apj}
\bibliography{bod.bib}

\clearpage

\end{document}